\documentclass[11pt]{article}
\usepackage{moriond,epsfig}

\newcommand{\vev}[1]{\left\langle#1\right\rangle}
\newcommand{\bfv}{\mathbf v}
\newcommand{\bfr}{\mathbf r}
\newcommand{\bfk}{\mathbf k}
\newcommand{\gtsima}{$ \buildrel > \over \sim \,$}
\newcommand{\ltsima}{$ \buildrel < \over \sim \,$}
\newcommand{\simgt}{\lower.5ex\hbox{\gtsima}}
\newcommand{\simlt}{\lower.5ex\hbox{\ltsima}}

\newcommand{\hmpc}{h^{-1}\,{\rm Mpc}}
\newcommand{\kms}{\ifmmode\,{\rm km}\,{\rm s}^{-1}\else km$\,$s$^{-1}$\fi}
\newcommand{\degs}{\ifmmode^\circ\else$^\circ$\fi}

\def\be{\begin{equation}}
\def\ee{\end{equation}}
\def\bea{\begin{eqnarray}}
\def\eea{\end{eqnarray}}

\begin{document}
\vspace*{4cm}
\title{COSMIC VELOCITIES 2000: A REVIEW \\
\medskip
{\rm To Appear in {\em Proceedings of the XXXVth
Rencontres de Moriond: \\
Energy Densities in the
Universe}}
}

\author{ J.A. WILLICK }

\address{Department of Physics, Stanford University, \\
Stanford, CA 94305-4060, USA }

\maketitle\abstracts{I review the status of cosmic velocity analysis
as of January 2000, with an emphasis on two key questions: (1) What is the
scale of the largest bulk flows in the universe? and (2) What is
the value of $\beta\equiv\Omega_m^{0.6}/b$ indicated by
cosmic velocities, and what does this tell us about $\Omega_m$
itself? These are the most important issues for cosmic flow
analysis, and each has been controversial in recent years.
I argue that a preponderance of the evidence at present
argues against very large scale ($\simgt 100\,\hmpc$) bulk flows,
and favors $\beta \simeq 0.4$--$0.5,$ corresponding to
a low-density ($\Omega_m \simeq 0.2$--$0.3$) universe.
}

\section{Background}

The study of cosmic flows emerged as a distinct subfield
of cosmology in the late 1970s, spurred by the discovery
of the Tully-Fisher (TF) 
and Faber-Jackson relations for spiral and elliptical galaxies,
respectively. With these empirical correlations between galaxy luminosity
and internal velocity as distance indicators,
one could measure redshift-independent
distances for galaxies out to tens or even hundreds
of megaparsecs. Both the
TF relation and the successors of Faber-Jackson---the $D_n$-$\sigma$
and Fundamental Plane (FP) relations---yield distances with $\sim 15$--$20\%$
accuracy. Although hopes for further improving the accuracy of TF
and FP have proved unfounded, these distance indicators
have remained the workhorses of cosmic velocity analysis
for two decades. 
Only in recent years have newer and more accurate distance
indicator methods---in particular Type Ia Supernovae
and Surface Brightness Fluctuations---begun to complement (not supplant)
TF and FP, as discussed further below.

The early scientific emphasis in flow studies was on determining the
amplitude of Virgo infall (e.g., Tonry \& Davis 1981; Aaronson et al.\ 1982).
The Virgo cluster and its environs
were then thought to dominate the local flow field. 
It has since been
recognized that the local velocity field is more complex.
There are a number of nearby attractors and voids, 
as well as the tidal effect of distant mass concentrations.
The most sophisticated models of the local velocity field
now use gravitational instability theory 
to predict peculiar velocities on the basis of the galaxy
density field observed in redshift surveys. The redshift
survey most widely used for this purpose is the IRAS redshift
survey, both in its older, 1.2 Jy (Fisher et al.\ 1995)
and newer PSCz (Saunders et al.\ 2000) incarnations. If one
assumes IRAS galaxies trace mass and adopts the approximations
of linear theory, comparison of predicted velocities with the
observed ones constrains the parameter $\beta_I=\Omega_m^{0.6}/b_I.$
Here, $b_I$ 
is the {\em biasing parameter\/} for IRAS galaxies,
a measure of whether IRAS galaxies are more ($b_I>1$) or
less ($b_I<1$) clustered than mass. (The subscript is needed
because different redshift samples have different clustering
amplitudes, and thus different biasing parameters.)

Measurement of $\beta_I$ has thus become one of the main
thrusts of cosmic flow analysis in the 1990s.
Early in the decade, when there was a widespread theoretical prejudice
in favor of an $\Omega_m=1$ universe, it was thought 
peculiar velocities might be the key to proving it.
On large scales, the argument went, galaxies should
trace the dominant dark matter, 
and the large-scale peculiar
field should therefore reflect the underlying Einstein-de Sitter
nature of the universe. The earliest efforts in
this direction indeed seemed to bear out this suspicion,
finding $\beta_I \approx 1.3,$ suggestive of $\Omega_m=1$
(Dekel et al.\ 1993). More recently, however,
cosmic flow-based estimates of $\beta_I$ have often,
though not invariably, produced values in the $0.4$--$0.6$
range, consistent with a low-density
cosmology. In \S 3 I will summarize
recent work done on this problem, and explain
why I believe the low $\beta_I$ values
are more likely to be correct.

Another aim of cosmic flow studies
first arose serendipitously: efforts to detect bulk flow on very
large ($\simgt 100\hmpc$) scales. This work was propelled
by the discovery, in 1987, of a bulk flow stretching
across the sky by the ``7-Samurai'' (7S) group 
(Dressler et al.\ 1987).
Working with the newly discovered $D_n$-$\sigma$ relation,
the 7S found that elliptical galaxies out to
$\sim 3000\,\kms$ redshift exhibited fairly uniform 
Hubble expansion from the vantage point of the Local
Group (LG) barycenter. 
But the LG is known to move
at $630\,\kms$ with respect to the Cosmic Microwave
Background radiation (CMB), as indicated by the CMB
dipole anisotropy. Thus, the implication of the 7S
findings was that the ellipticals were streaming
at $\sim 600\,\kms$ relative to the CMB.
If there is any validity to Big Bang cosmology, though,
the CMB defines an absolute standard of rest---the ``cosmic
rest frame,'' as it were. 
The 7S finding thus inaugurated a long-standing
puzzle in cosmology: how can large-scale, coherent
bulk flows exist in a universe that seems to be so 
uniform on large scales?

The puzzle deepened in the following decade, 
with a number of groups
confirming a 7S-like bulk flow, and, in several cases,
finding that it continued to scales three or four times 
the 7S volume. 
The current controversy may be framed as ``what is the scale
of the largest bulk flow,'' or, equivalently,
as one of {\em convergence scale:} at what
distance are the galaxies within a spherical shell
finally at rest in the CMB frame? Some astronomers
(this author included) have been led to wonder whether such a
convergence scale existed; perhaps the CMB-defined
``cosmic rest frame'' was offset by $600\,\kms$ from
the frame in which uniform Hubble expansion is observed---a
conjecture which, if correct, would call into
question some of the fundamental tenets of cosmology.
Fortunately---at least if you like agreement between
theory and observation---it now appears that the
convergence to the CMB has been detected at a distance
of $\sim 50$--$60\hmpc.$ Or at least, that is what
I will argue in \S 2, when
I summarize results from newly completed surveys. 

\section{The Scale of the Largest Bulk Flows}

First we should ask, from the perspective of cosmology, 
Why are bulk flows interesting? 
In particular, what does their convergence scale tell us?

The answer lies in the sensitivity of bulk flows to
long-wavelength modes of the mass fluctuation power spectrum.
Mass conservation in the linear regime of
gravitational instability tells us that
\be
\nabla \cdot \bfv = -\Omega_m^{0.6} \delta,
\label{eq:vdens}
\ee
where $\bfv$ is the peculiar velocity vector and
$\delta$ is the mass density contrast.  
The corresponding equation in Fourier space is
$\bfk \cdot \bfv_k = -\Omega_m^{0.6} \delta_k,$
where the subscript denotes Fourier transform. 
Thus, $|\bfv_k| \propto \delta_k/k,$ which is to
say, long-wavelength perturbations (small $k$)
have a larger impact on large-scale peculiar velocities than they
do on mass fluctuations. 

\begin{figure}[ht!]
\begin{center}
\includegraphics[scale=0.375]{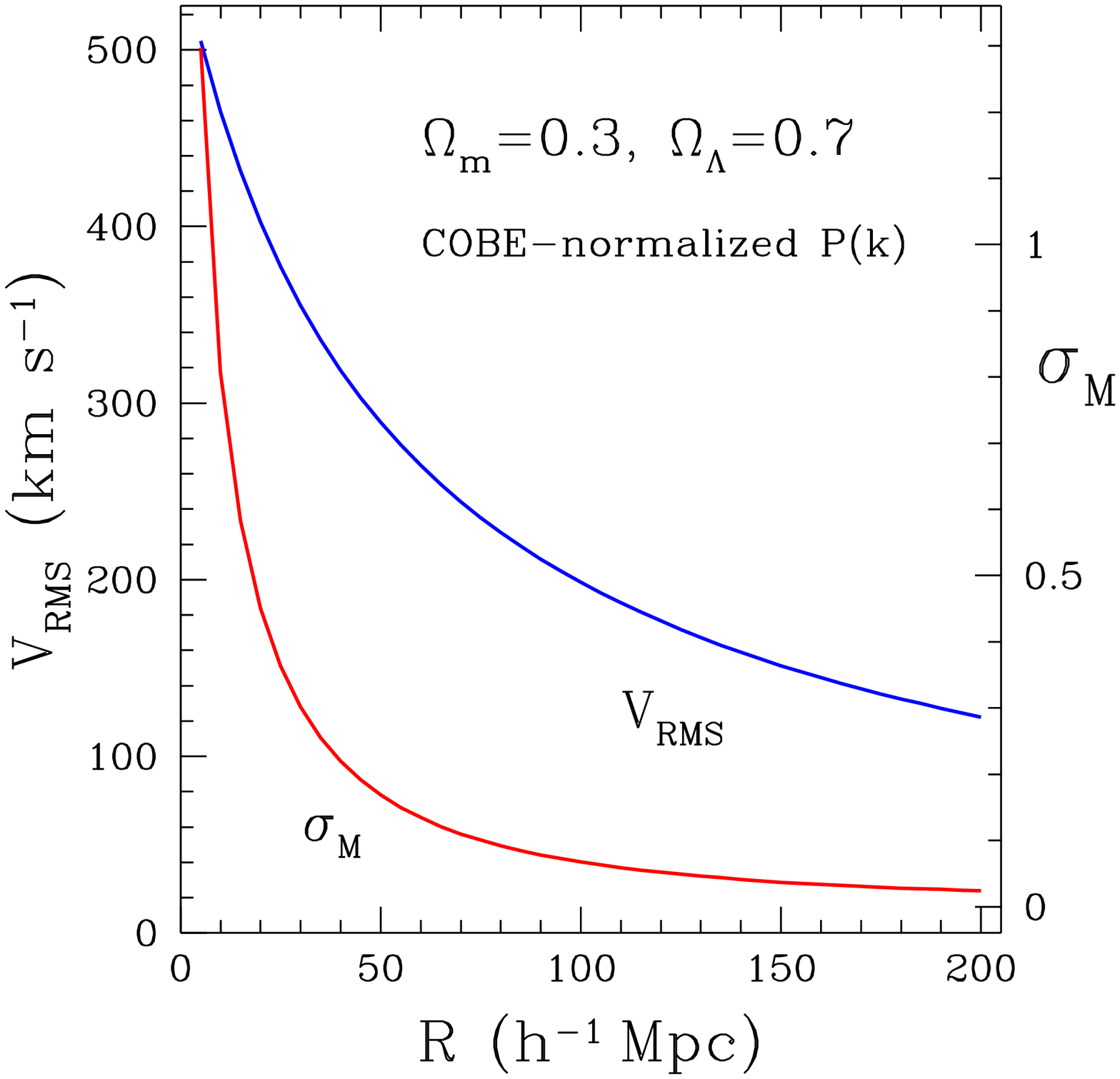}\hfill~
\includegraphics[scale=0.39]{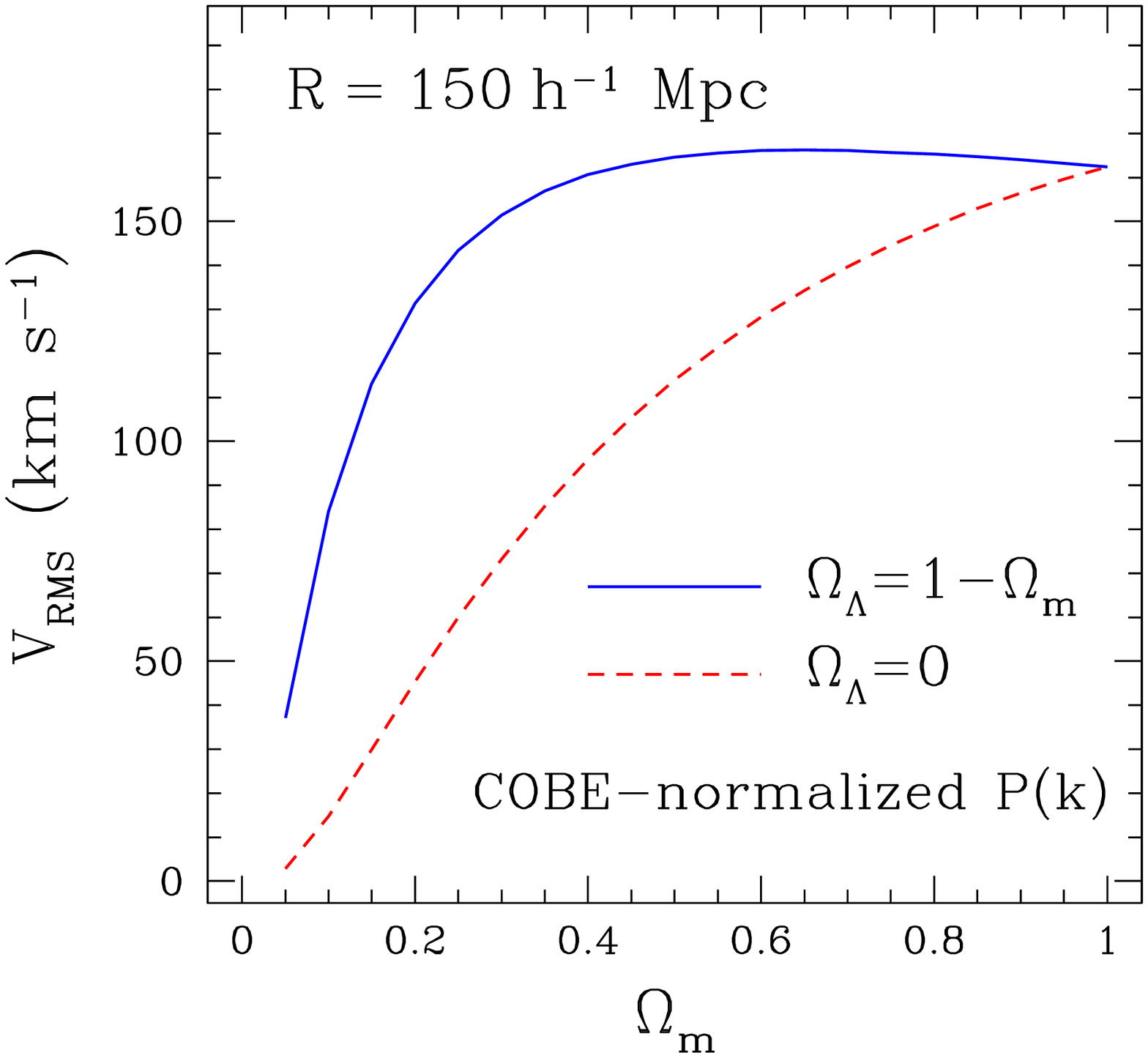}
\end{center}
\vspace{0.1cm}
\caption{{\small Left panel: Expected rms bulk velocity of a sphere
of radius $R,$ for an $\Omega_m=0.3,$ $\Omega_\Lambda=
0.7$ universe. Also shown is
the mass density fluctuation $\sigma_{_M}(R).$ Right panel:
rms bulk velocity of an $R=150\hmpc$ sphere
versus $\Omega_m$ for open and flat universes.}}
\label{fig:vrms}
\end{figure}

One can flesh out these ideas by calculating the
mean square bulk velocity on a scale $R:$
\be
\vev{v^2(R)} = \frac{\Omega_m^{1.2}}{2\pi^2}
\int_0^\infty P(k)\, \widetilde W^2(kR) \, dk \,,
\label{eq:Vrms}
\ee
where $P(k)$ is the mass fluctuation power spectrum
and $\widetilde W^2(kR)$ is the Fourier transform
of a top-hat window of radius $R.$ In Figure~\ref{fig:vrms}
the rms expected bulk velocity, $V_{rms}(R)=\sqrt{\vev{v^2(R)}},$
is plotted against $R$ (left panel) and $\Omega_m$ (right panel)
for COBE-normalized CDM power spectra. The left panel assumes
a canonical $\Omega_m=0.3,$ $\Omega_\Lambda=0.7$ universe.
For comparison, the rms density fluctuation $\sigma_{_M}(R)=
\vev{(\delta M)/M)^2(R)}$ is also plotted. Note that $\sigma_{_M}$
drops much more rapidly with scale than does $V_{rms}.$
This is a result of the sensitivity of large-scale bulk flow 
to long-wavelength modes of the power spectrum. An observational
consequence is that while redshift surveys have difficulty  
probing the power spectrum on scales $\simgt 100\hmpc,$ bulk
flow studies can in principle do so. 

Closer inspection of the left panel also
shows why large-amplitude ($V\simgt 500\,\kms$), large-scale
$R \simgt 60\hmpc$ are potentially problematic for standard
structure-formation scenarios. In the CDM-type models shown
here, there simply isn't enough large-scale power to drive
such flows. Stated another way, the CDM universe approaches 
homogeneity sufficiently rapidly with increasing scale
that the coherence scale of bulk flows should be a few
tens of megaparsecs at most. Moreover, the right panel
shows, perhaps counterintuitively, that 
boosting the matter density doesn't help. (This is a consequence
of imposing COBE-normalization; normalizing the power spectrum
to the cluster abundance does not substantially change
our conclusions.)


However, high-amplitude bulk motions on scales $\simgt 100\hmpc$
are just what were found by three surveys
conducted in the early and mid 1990s. The first of these, and the
one which has achieved the greatest notoriety, was the Brightest
Cluster Galaxy (BCG) survey by Lauer \& Postman (1994; LP). LP used the
BCGs as standard candles, thus obtaining distances to over 100 Abell clusters
out to $15,000\,\kms$ redshift, and found them to be
coherently moving relative to the CMB at $\sim 700\,\kms.$ 
The combination of large scale and high amplitude places the LP
result far above the expected $V_{rms}$ values in Figure~\ref{fig:vrms}.

More recently, two large surveys of cluster galaxies appeared
to confirm the scale and amplitude (but not the direction) of
the LP result. The Streaming Motions of Abell Clusters (SMAC)
survey of Hudson et al.\ (1999) used the FP relation to
measure cluster ellipticals to about the same depth
as the LP survey, and found a $600\,\kms$ flow, in roughly
the same direction as the that found by 7S a decade earlier.
And in a Tully-Fisher survey of cluster galaxies in
a shell between $\sim 9000$ and $13,000\,\kms,$ I (Willick 1999,
LP10K) found
a streaming motion of about $700\,\kms$ in roughly the
same direction as SMAC. The above results are summarized
in Table I, along with another bulk flow measurement based on 
more nearby galaxies from the Mark III Catalog of TF 
and $D_n$-$\sigma$ data (Willick et al.\ 1997),
as measured by the POTENT algorithm (on which more
in \S~3).

\begin{center}
\centerline{{\sc Table I. Recent Bulk Flow Measurements}}
\begin{tabular}{l c c | l}\hline\hline
Survey & $R$ ($\kms$) & $V_B$ ($\kms$) & Comments \\ \hline
Lauer-Postman (LP) & 15000 & 700 &  BCG \\
Willick (LP10K) &  12000  &  700 & TF  \\
Hudson et al.\ (SMAC)  &  14000  &  600 & FP  \\
Dekel et al.\ (POTENT) &  6000   & 350 & TF+$D_n$-$\sigma$ \\ \hline
\end{tabular}
\end{center}
\medskip

The four results above argue for large bulk flows, but are
not fully consistent. The SMAC, LP10K, and POTENT/Mark III flows
agree in direction, but the LP flow is nearly orthogonal.
Also, the smaller amplitude of the smaller scale POTENT/Mark III 
measurement is puzzling; one would expect bulk flow amplitude to
diminish monotonicaly with scale (Figure~\ref{fig:vrms}). Even
without evidence to the contrary, then, the above results are
less than convincing.

More importantly, however, 1999 saw the announcements of
new survey results that contradict the findings of Table 1.
The nearby Type Ia Supernova (SN Ia) data have accumulated, and
as reported by Riess (2000), the sample of SN Ia distances
within $\sim 10,000\,\kms$ redshift show no evidence
for bulk flow. This is quite important, because the SN Ia
data are of a fundamentally different nature than the other
distance indicators employed in cosmic flow studies. Also,
as shown in the cosmological context, SN Ia have small
scatter, $\sim 0.15$ mag. The EFAR FP survey of Colless et
al.\ (2000) similarly finds no bulk motion on a
large scale, and in particular is inconsistent
with LP at better than 99\% confidence. Similar
results have been obtained by Dale, Giovanelli, and
coworkers from their extensive TF surveys (as summarized
by Dale \& Giovanelli 2000). The key findings of
these surveys, plus that of the Shellflow survey, to
which I turn next, are summarized in Table~2.

\medskip
\begin{center}
\centerline{{\sc Table II. More Recent Bulk Flow Measurements}}
\begin{tabular}{l r c | l}\hline\hline
Survey & $R$ ($\kms$) & $V_B$ ($\kms$) & Comments \\ \hline
Riess et al.\ & $10000$ & $\approx 0$ &  SN Ia \\
Courteau et al.\ (SHELLFLOW) &  $6000$  &  $\approx 0$ & TF  \\
Colless et al.\ (EFAR)  &  $\sim 10000$  &  $\approx 0$ & FP  \\
Dale \& Giovanelli (SFI) &  $6000$   & $\approx 0$ & TF \\
Dale \& Giovanelli (SCI/SCII) &  $\sim 15000$   & $\approx 0$ & TF \\ \hline
\end{tabular}
\end{center}
\medskip

A group of us who were active in cosmic velocity
measurement and analysis (M.\ Strauss, S.\ Courteau,
M.\ Postman, D.\ Schlegel, and myself) realized
in 1995 that a critical issue had
not been properly addressed in the then-extant
Tully-Fisher surveys: the need for extremely uniform
data across the sky. We proposed for and were granted extensive
NOAO time for the Shellflow project, a TF survey of
a shell of 300 galaxies between $5000$ and $7000\,\kms$ redshift.  
By using identical observational setups from northern
and southern hemisphere NOAO telescopes we ensured
that data nonuniformity could not produce spurious
peculiar velocities. 

\begin{figure}[h!]
\begin{center}
\includegraphics[scale=0.60]{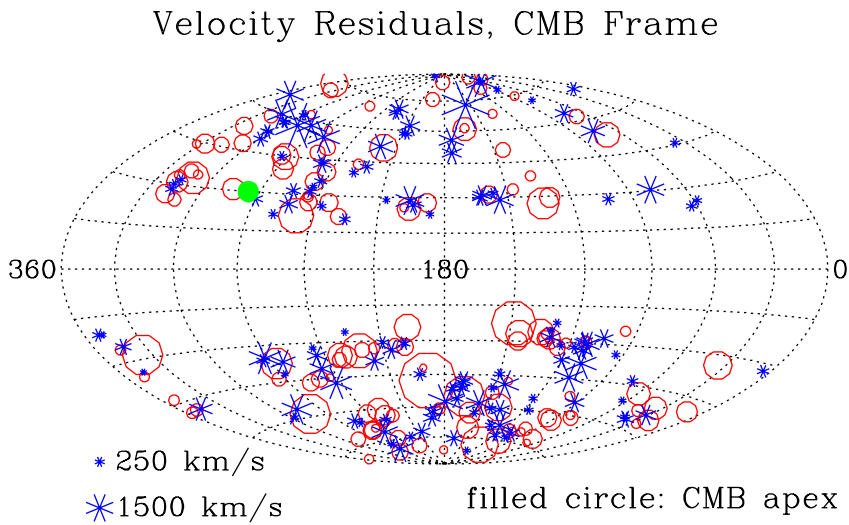}
\end{center}
\caption{{\small 
Apparent peculiar velocity residuals (TF residuals converted
into velocities) of 
Shellflow galaxies for a pure Hubble flow fit. 
Point size is proportional to the velocity amplitude,
with fiducial values indicated at the lower left.
Circles and asterisks represent
inflowing and outflowing objects, respectively.}}
\label{fig:cmbresid}
\end{figure}

In a recent paper (Courteau et al.\ 2000) we reported
our main result: the bulk flow of the shell
centered at $6000\,\kms$ is $70^{+100}_{-70}\,\kms,$
i.e., is consistent with being at rest in the CMB
frame. The residuals with respect to a fit
that assumes pure Hubble flow in the CMB frame
are shown in Figure~\ref{fig:cmbresid}. The points are
everywhere consistent with being due to TF scatter,
not coherent peculiar velocities. Inflowing and
outflowing points are well mixed at all positions,
indicating the absence of coherent motions.

Having worked with the Shellflow data myself, I am
confident that it is of very good quality, and am
convinced that systematic errors have a very small effect, if any
on our results. I therefore find the conclusion of
insignificant bulk flow highly persuasive. 
The Shellflow results refer to a specific scale,
namely, a $60\hmpc$ sphere. They do not directly
test the LP, SMAC, and LP10K results of Table 1.
However, because abundant evidence indicates
that the universe approaches homogeneity monotonically
with increasing scale size, I believe that
the Shellflow result, if correct, is {\em physically\/}
inconsistent with the LP, SMAC, and LP10K findings,
and that the latter are therefore not to be taken at face value.
 
One should not, however, judge the LP, SMAC,
and LP10K authors harshly (and I have an obvious
reason for hoping you don't!). Measuring distortions
of the Hubble expansion that are at the level of a few
percent is a challenging task, given
the limitations of the distance indicator 
techniques we work with. 
False detections, if that is what they are, will
most likely be seen in hindsight as inevitable
products of initial efforts at a very
difficult measurement.

\section{The Value of $\beta_I$}

In considering the scientific implications of large-scale
bulk flow surveys, I made no mention of comparison with
redshift surveys. That is because the full-sky redshift
surveys we have are not reliable enough, at distances $\simgt
150\hmpc,$ to predict what the bulk flows {\em should\/}
be on such scales. 
The situation changes when we talk about the 
velocity field within $\sim 50\hmpc.$ At these
distances both the peculiar velocity and
galaxy density fields are mapped with enough
accuracy to do a detailed comparsion. The goals
of this comparison are (1) to verify that the
two maps are compatible with the gravitational
instability paradigm, and (2) assuming they are,
to measure $\beta_I=\Omega_m^{0.6}/b_I,$ and
if possible to go further and measure $\Omega_m$ itself.\footnote{It
will be assumed in what follows that, unless otherwise specified,
the redshift survey in question is one of IRAS galaxies.}

The theoretical bases of the 
comparison are either of two forms
of the linear velocity-density relation. One is the
differential form,
Eq.~(\ref{eq:vdens}), which for comparison of observables is written as
\be
\nabla\cdot\bfv(\bfr) = - \beta_I\delta_{I,g}(\bfr) \,,
\label{eq:viras}
\ee
where $\delta_{I,g}(\bfr)$ is the galaxy overdensity as determined
by an IRAS redshift survey.
The other is the
corresponding integral form,
\be
\bfv(\bfr) = \frac{\beta_I}{4\pi} \int d^3\bfr'\,\frac{\delta_{I,g}(\bfr')
(\bfr' - \bfr)}{\left|\bfr' - \bfr\right|^3} \,.
\label{eq:vpdelta}
\ee
In both Eqs.~(\ref{eq:viras}) and~(\ref{eq:vpdelta}), the
spatial positions $\bfr$ are assumed to be measured in
velocity units---i.e., they are distances in units of
the Hubble velocity. In this way, the Hubble constant
itself is removed from the analysis, which is (needless to
say) a useful simplification.

Although the two forms of the velocity-density relation
are equivalent, they lead to two rather different analytical
approaches to measuring $\beta_I.$ Because the controversy
between the ``high'' and the ``low'' (see \S 1) values of
$\beta_I$ appears to revolve around this distinction,
it is worth taking a moment to understand it.  
To apply Eq.~(\ref{eq:viras}), one needs to map out
the {\em three-dimensional\/} velocity field $\bfv(\bfr),$
differentiate it, and finally compare it to
the galaxy density field to determine $\beta_I.$
Because only the radial component of $\bfv$ is observable,
one first needs an {\em ansatz\/} for ``three-dimensionalizing''
the inherently one-dimensional velocity data. An
elegant approach to this problem was developed by
Bertschinger \& Dekel (1989; see Dekel 1994 for a review),
who argued that the large-scale velocity field
should be irrotational and thus expressible as
the gradient of a potential function, which could itself 
be computed by integrating the observed, radial velocities
along rays. This algorithm,
known as POTENT, thus produces a 3D
velocity field, smoothed on a rather large
(typically $12\hmpc$) scale, which can be differentiated
and then used in Eq.~(\ref{eq:viras}). 

Eq.~(\ref{eq:vpdelta}) suggests a different approach. Rather
than heavily processing the velocity data, one carries
out the indicated
integration using the redshift survey data. One thus
obtains a predicted peculiar velocity field as a
function of $\beta_I,$ of
which only the radial component, $u(\bfr)=\bfv(\bfr)\cdot\bfr/r,$
is used in the subsequent analysis. The predicted $u(\bfr)$
is compared with the observed radial peculiar velocities
from the TF (FP, etc.) data sets; the final estimate of
$\beta_I$ is that which yields the closest match
predictions and observations.

The two approaches to measuring $\beta_I$ are known as
the {\em density-density\/} (d-d) and
{\em velocity-velocity\/} (v-v) comparisons. It is notable
that d-d comparisons, all done using POTENT
to reconstruct the 3D velocity field, have consistently produced
values of $\beta_I$ consistent with unity---and thus,
to the extent $b_I$ is itself not so different from one,
implicit estimates of $\Omega_m$ near unity as well. 
The 1993 paper by Dekel et al.\ was already mentioned
(see \S 1). A more recent application of POTENT, using improved
peculiar velocity data, was that of Sigad et al.\ (1998),
who found $\beta_I=0.89 \pm 0.1.$ 

Since about 1995, several v-v alternatives to the POTENT approach
have been developed for measuring $\beta_I.$ They have
differed in the way in which the IRAS (or other) redshift data
are used to predict peculiar velocities, and in the
way the predicted and observed peculiar velocities
are compared. But as v-v methods they have more
in common with one another than with POTENT;
in particular, the heavy computational work
is done with the redshift data, while the TF (FP, etc.)
data are used only in a limited, statistical sense.
These methods include the Least Action Principle
approach of Shaya, Tully, \& Peebles (1995), who obtain
$\beta_I \approx 0.35 \pm 0.1$), the VELMOD method of
Willick et al. (1997) and Willick \& Strauss (1998),
who obtain $\beta_I=0.5 \pm 0.05,$ and the ITF method
(Davis, Nusser, \& Willick 1996) which, applied to
the Type Ia Supernova data produced a value
of $\beta_I=0.4\pm 0.1.$

\begin{figure}[ht!]
\vspace{1cm}
\begin{center}
\includegraphics[scale=0.39]{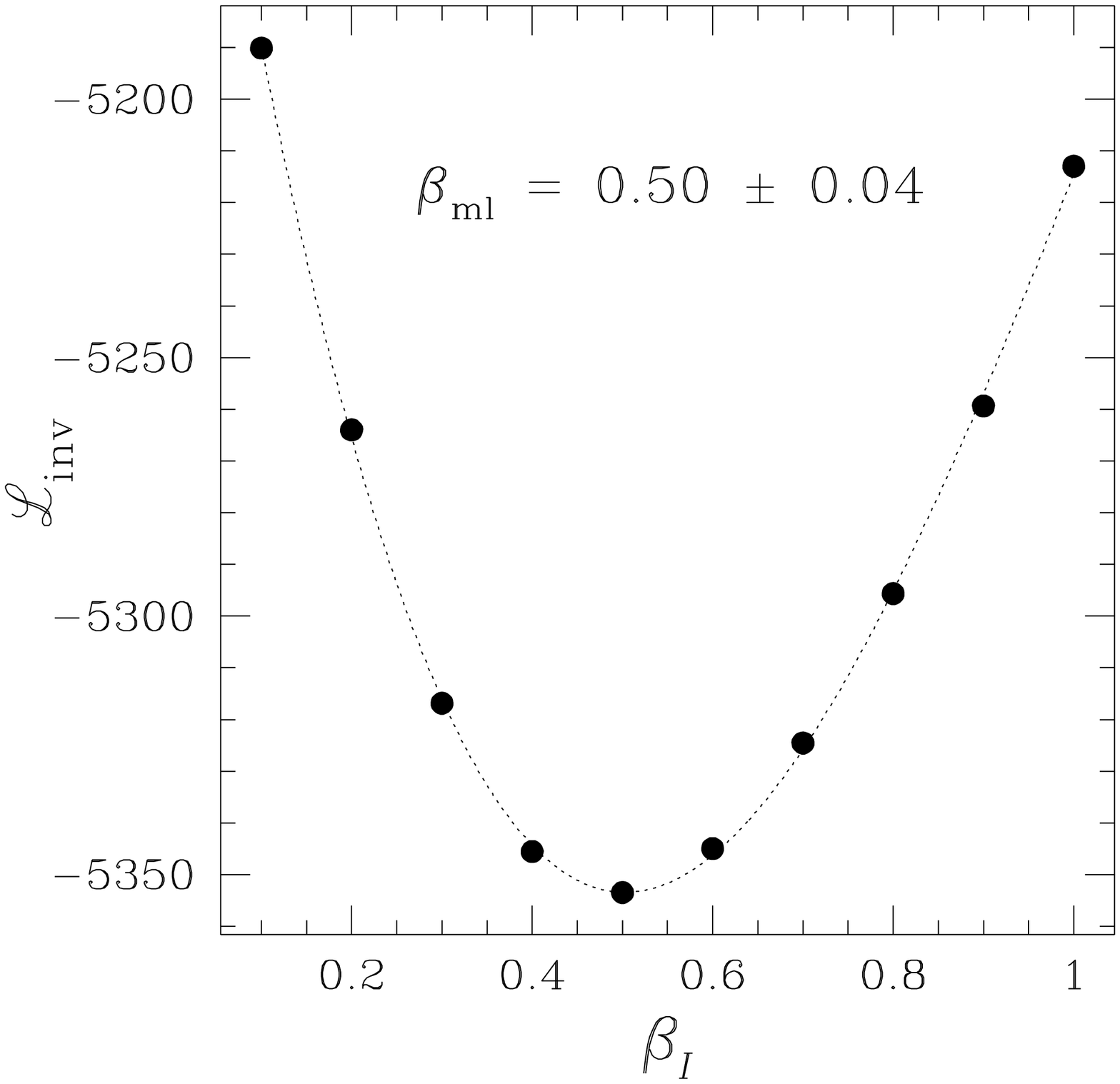}\hfill~
\includegraphics[scale=0.47]{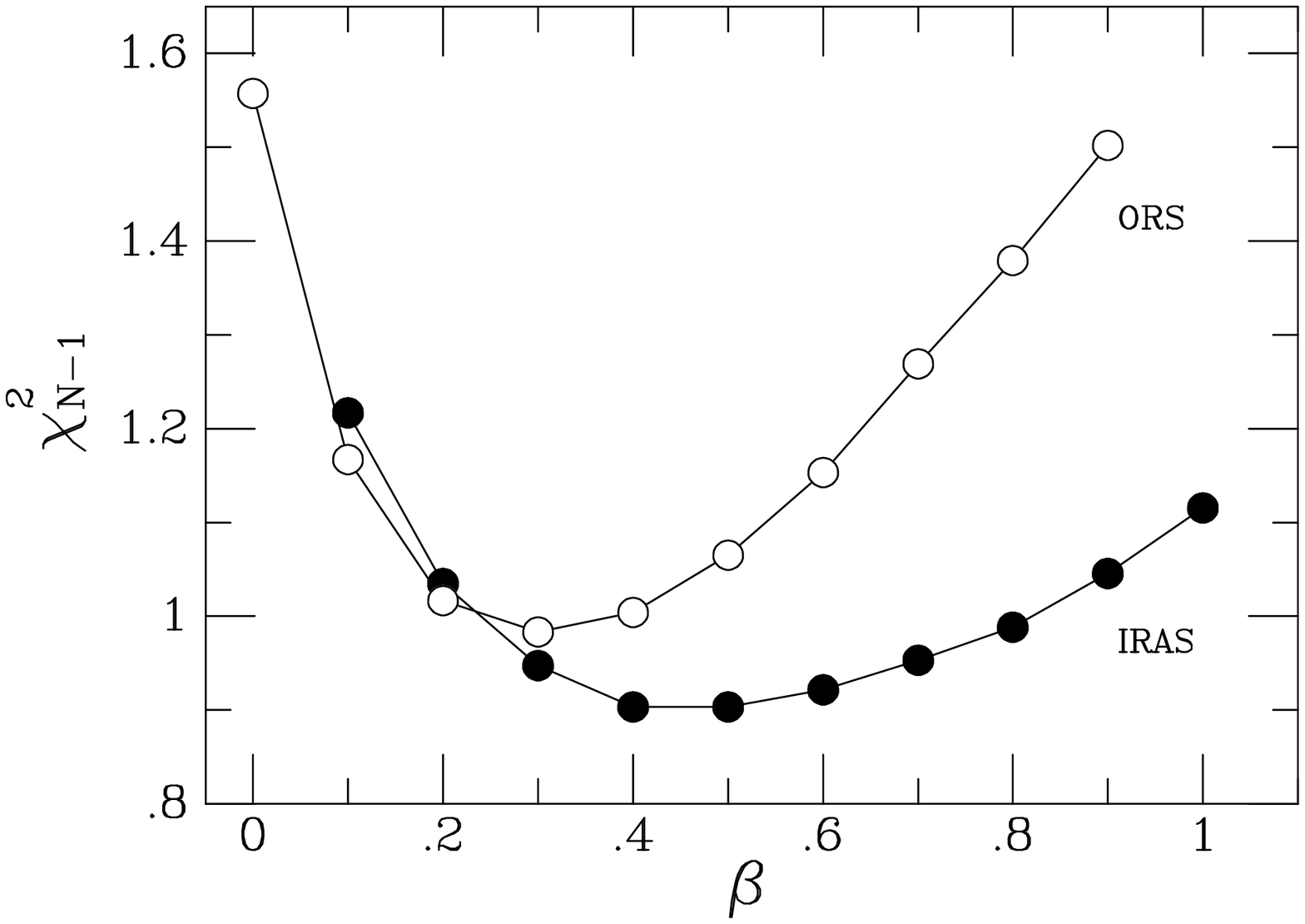}
\end{center}
\vspace{0.3cm}
\caption{{\small Left panel: VELMOD likelihood statistic ${\cal L}$
versus $\beta_I,$ for the IRAS-predicted velocity field
and the Mark III Catalog Tully-Fisher data.
Adapted from Willick \& Strauss (1998).
Right panel:
Reduced $\chi^2$ versus $\beta$ for the SBF velocity 
sample compared with IRAS and Optical Redshift Survey
(ORS) predictions. The best fit
for IRAS is $\beta_I=0.44 \pm 0.08.$ Adapted
from Blakeslee et al.\ (2000).}}
\label{fig:velmod}
\end{figure}

Figure~\ref{fig:velmod} shows representative results from
two v-v analyses. The left panel shows the VELMOD
results from Willick \& Strauss (1998). The statistic
plotted is ${\cal L}=-2\ln P,$ where $P$ is the probability
of observing the TF data given the IRAS velocity model.
(The calculation is done for the ``inverse'' TF relation,
which is immune to selection biases.) The minimum of the
curve occurs at the maximum likelihood value of
$\beta_I,$ and its curvature yields the $1\,\sigma$ error,
as indicated on the plot.

The right panel shows the first use of the new Surface
Brightness Fluctuation (SBF) data set in $\beta$-measurement,
as reported by Blakeslee et al. (2000). As was stated
above in connection with SN Ia data, it is essential
to carry out these experiments with independent distance
indicators, and the SBF method is quite distinct from
Tully-Fisher. Moreover, like SN Ia, SBF distances 
are on average about twice as accurate as TF distances,
and have the additional advantage of having well-determined
errors. In Figure~\ref{fig:velmod}, the SBF data are
compared with velocities predicted by IRAS and the 
Optical Redshift Survey (ORS). The minimum $\chi^2$ is achieved
for $\beta_I=0.44\pm 0.08,$ consistent with the VELMOD
TF results. (The ORS galaxies are more clustered than
IRAS galaxies, and therefore yield a lower $\beta$ value.)

A truly convincing explanation for this discrepancy between
the v-v and d-d $\beta$ values has not
yet been found. It seems to me, however, that the v-v
comparison is a more robust procedure. In it, the intensive
data manipulation is done on the redshift survey
data, which is by its nature more reliable than
the distance indicator data (redshifts errors
are fractionally very small; redshift-independent
distance errors are always $\sim 20\%$).
In the d-d comparison, it is the data set 
with much larger scatter
and non-Gaussian errors that is subjected to 
intensive manipulation. In particular, the d-d comparison
requires that noisy data first be smoothed, then integrated,
and then differentiated in three dimensions. There is ample opportunity
in this procedure for errors to propagate. The
same noisy data in the v-v studies are left
virtually untouched, save for the statistical
comparison with their predicted values. This
procedure is far more stable. For this reason I
consider the low $\beta$ values derived from
the v-v analyses to be more reliable. 

\section{Summary}

I have argued that the two major controversies in cosmic flow
analysis have been
largely resolved in the last few years. 
With regard to bulk flows, most recent surveys show
convergence to the CMB frame by a distance of
$\sim 60\hmpc.$  A corollary is that the
observed bulk motions do not require more large-scale power  
than is provided by COBE-normalized CDM density fluctuation spectra.
With regard to the value of $\beta_I=\Omega_m^{0.6}/b_I,$ a
number of independent analyses now suggest a low value,
$\beta_I \simeq 0.4$--$0.5.$ If the IRAS galaxies are
nearly unbiased with respect to mass, a reasonable
if not airtight hypothesis, these $\beta$-values imply
a density parameter $\Omega_m \simeq 0.2$--$0.3.$

A word of caution is in order, however. The above conclusions
represent a consensus view, 
not a unanimous one. The bulk flow
detections listed in the first three
rows of Table~1 have not been in any sense ``refuted,''
which is to say as far as we know there is nothing wrong
with the data. 
New surveys, such as FP200 
(see http://astro.uwaterloo.ca/$\sim$mjhudson/fp200 for details)
will, it is hoped, settle the issue definitively. 
Similarly, while a majority of recent velocity-density comparisons
favor low $\beta_I,$ the reason for the
discrepant POTENT result, $\beta_I \approx 0.9,$
is not well understood. Tests of methods such as 
POTENT and VELMOD using N-body simulations are under
way, and may clarify things. Moreover, the coming decade
will bring much larger TF data sets
from the DENIS and 2MASS infrared surveys. As always,
these new data sets, if they live up to their promise, are our
best hope for putting any remaining
contoversy to rest. 

\section*{Acknowledgments}
I am grateful to my
collaborators on the Shellflow project, St\'ephane Courteau,
Michael Strauss, Marc Postman, and David Schlegel, 
as well as to Michael Hudson, Marc Davis, Avishai Dekel,
John Tonry and Alan Dressler for enlightening
discussions over the last several years. Special thanks go to St\'ephane
Courteau for organizing the highly successful Cosmic Flows 99
conference in Victoria, B.C. last summer.
My research is supported by
a Cottrell Scholarship of Research Corporation,
NSF grant AST96-17188, and a Terman Fellowship from Stanford
University.

\end{document}